# Caging-Pnictogen-Induced Superconductivity in Skutterudites IrX$_3$ (X = As, P)

Cuiying Pei[a†], Tianping Ying[b,c*†], Qinghua Zhang[c†], Xianxin Wu[d,e], Tongxu Yu[f], Yi Zhao[a], Lingling Gao[a], Changhua Li[a], Weizheng Cao[a], Qing Zhang[a,g], Andreas P. Schnyder[d], Lin Gu[c], Xiaolong Chen[c], Hideo Hosono[b*], Yanpeng Qi[a,g,h*]

[a]School of Physical Science and Technology, ShanghaiTech University, 393 Middle Huaxia Road, Shanghai 201210, China

[b]Materials Research Center for Element Strategy, Tokyo Institute of Technology, 4259 Nagatsuta, Midori-ku, Yokohama 226-8503, Japan

[c]Beijing National Laboratory for Condensed Matter Physics, Institute of Physics, Chinese Academy of Sciences, Beijing 100190, China

[d]Max-Planck-Institut für Festkörperforschung, Heisenbergstrasse 1, D-70569 Stuttgart, Germany

[e]CAS Key Laboratory of Theoretical Physics, Institute of Theoretical Physics, Chinese Academy of Sciences, Beijing 100190, China

[f]Gusu Laboratory of Materials, Jiangsu 215123, China

[g]Shanghai Key Laboratory of High-resolution Electron Microscopy, ShanghaiTech University, Shanghai 201210, China

[h]ShanghaiTech Laboratory for Topological Physics, ShanghaiTech University, Shanghai 201210, China

*E-mail: qiyp@shanghaitech.edu.cn; ying@iphy.ac.cn; hosono@mces.titech.ac.jp

†These authors contributed equally

**ABSTRACT:**

Here we report on a new kind of compound, X$_\delta$Ir$_4$X$_{12-\delta}$ (X = P, As), the first hole-doped skutterudites superconductor. We provide atomic resolution images of the caging As atoms using scanning transmission electron microscopy (STEM). By inserting As atoms into the caged structure under a high pressure, superconductivity emerges with a maximum transition temperature ($T_c$) of 4.4 K (4.8 K) in IrAs$_3$ (IrP$_3$). In contrast to all of the electron-doped skutterudites, the electronic states around the Fermi level in X$_\delta$Ir$_4$X$_{12-\delta}$ are dominated by the caged X atom, which can be described by a simple body-centered tight-binding model, implying a distinct paring mechanism. Our density functional theory (DFT) calculations reveal an intimate relationship between the pressure-dependent local-phonon mode and the enhancement of $T_c$. The discovery of $X_\delta Ir_4 X_{12-\delta}$ provides an arena to investigate the uncharted territory of hole-doped skutterudites, and the method proposed here represents a new strategy of carrier doping in caged structures, without introducing extra elements.

## INTRODUCTION

Cage-structured compounds are composed of covalently bonded cage-forming frameworks, which can accommodate guest atoms in their interior. The incorporation of guest atoms significantly influences the physical properties of the host materials: carrier doping alters the Fermi level ($E_F$), thereby modifying the mobility and band structure around $E_F$. At the same time, the caged guest atoms induce some low-energy vibrational modes, a unique feature of the caged materials, which can effectively lower the thermal conductivity[1-3]. Given the high tunability of these structures and their distinct charge carrier and phonon behavior, multifarious physical phenomena have been discovered in this class of caged materials, including thermoelectricity[4], superconductivity[5], and heavy-fermion behavior[6]. Representative and well-studied examples of caged materials are zeolites, clathrates, pyrochlores, and skutterudites[7-9].

The mineral skutterudites have been extensively investigated both in their primary constitution and cage-filled form. The cage of the skutterudites can easily host a number of alkali metals, alkaline earth metals, or lanthanides, which donate electrons to the host and thereby significantly alter the physical properties[10]. Moreover, the strong Coulomb attraction between the caged cation (alkali/alkali-earth metals) and the surrounding arsenide substantially enhances electronic correlations, which leads to complicated band features near $E_F$ and induces numerous interesting phenomena including superconductivity. Thus far, all the superconductors discovered in the skutterudites class have been limited to electron-doped (*n*-type) (Table 1). Hole-doping (*p*-type doping), however, has not been feasible so far through conventional synthesis routes, due to the strong Coulomb repulsion between the *p*-dopant and the structural P/As, and due to the covalent nature (i. e. saturation and directionality) of the bonds. Consequently, phenomena induced by hole doping, such as superconductivity, could not be explored in the skutterudites.

**Table 1. A list of superconducting skutterudites.**

| Compound | $a$ / Å | $T_c(P)$ / K(GPa) | Type | Ref. | Compound | $a$ / Å | $T_c(P)$ / K(GPa) | Type | Ref. |
|---|---|---|---|---|---|---|---|---|---|
| **IrAs$_3$** | **7.7397** | **4.4(93.7)** | **p** | **a** | **IrP$_3$** | **7.4286** | **4.8(98.7)** | **p** | **a** |
| Ba$_{0.85}$Ir$_4$As$_{12}$ | 8.5605 | 4.8(0) | n | 5 | Ba$_{0.89}$Ir$_4$P$_{12}$ | 8.1071 | 5.6(0) | n | 5 |
| BaPt$_4$Ge$_{12}$ | 8.6838 | 5.35(0) | n | 11-12 | SrPt$_4$Ge$_{12}$ | 8.6509 | 5.1(0) | n | 11-12 |
| YFe$_4$P$_{12}$ | 7.789 | 7(0) | n | 13-14 | YRu$_4$P$_{12}$ | 8.0298 | 8.5(0) | n | 13, 15 |
| YOs$_4$P$_{12}$ | 8.0615 | 3(0) | n | 13, 16 | LaOs$_4$P$_{12}$ | 8.0844 | 1.8(0) | n | 13, 17 |
| LaOs$_4$Sb$_{12}$ | 9.3055 | 0.74(0) | n | 13, 18 | LaOs$_4$As$_{12}$ | 8.5437 | 3.2(0) | n | 19 |
| LaRu$_4$P$_{12}$ | 8.0605 | 7.2(0) | n | 13, 17 | LaRu$_4$Sb$_{12}$ | 9.2781 | 3.58(0) | n | 20 |
| LaRu$_4$As$_{12}$ | 8.5081 | 10.3(0) | n | 21 | LaRh$_4$P$_{12}$ | 8.0788 | 16.5(0) | n | 22 |
| LaFe$_4$P$_{12}$ | 7.8316 | 4.1(0) | n | 13, 17 | LaPt$_4$Ge$_{12}$ | 8.6235 | 8.3(0) | n | 12 |
| PrRu$_4$P$_{12}$ | 8.0493 | 2(12) | n | 23 | PrRu$_4$Sb$_{12}$ | 9.2648 | 1.3(0) | n | 20 |
| PrRu$_4$As$_{12}$ | 8.4963 | 2.4(0) | n | 13, 24 | PrOs$_4$Sb$_{12}$ | 9.3031 | 1.85(0) | n | 25-26 |
| PrPt$_4$Ge$_{12}$ | 8.6111 | 7.9(0) | n | 12 | ThPt$_4$Ge$_{12}$ | 8.5931 | 4.75(0) | n | 27 |

[a]this work

Herein, we report the realization of hole-doped skutterudites by self-insertion of As and P into the in-cage sites of IrX$_3$ (X = P, As) under high pressure. The successful insertion of the center As is substantiated by STEM images, X-ray diffraction (XRD), Hall measurements, and theoretical calculations. Accompanied with the atom insertion, superconductivity emerges with a continuous enhancement of $T_c$ up to 4.8 K. This structural alteration induced by the atom insertion results in much-simplified energy bands near the Fermi level, which are dominated by the $p_x$, $p_y$ and $p_z$ orbital contributions of the caging atom. This feature facilitates the accurate description of the newly generated bands by a simple *bcc* tight-binding model. Combined with phonon spectra simulation, we highlight the importance of the caging atom for the realization of superconductivity. The proposed hole doping scenario should be applicable to other caged materials to exploit more unconventional physical behavior.

**RESULTS AND DISCUSSION**

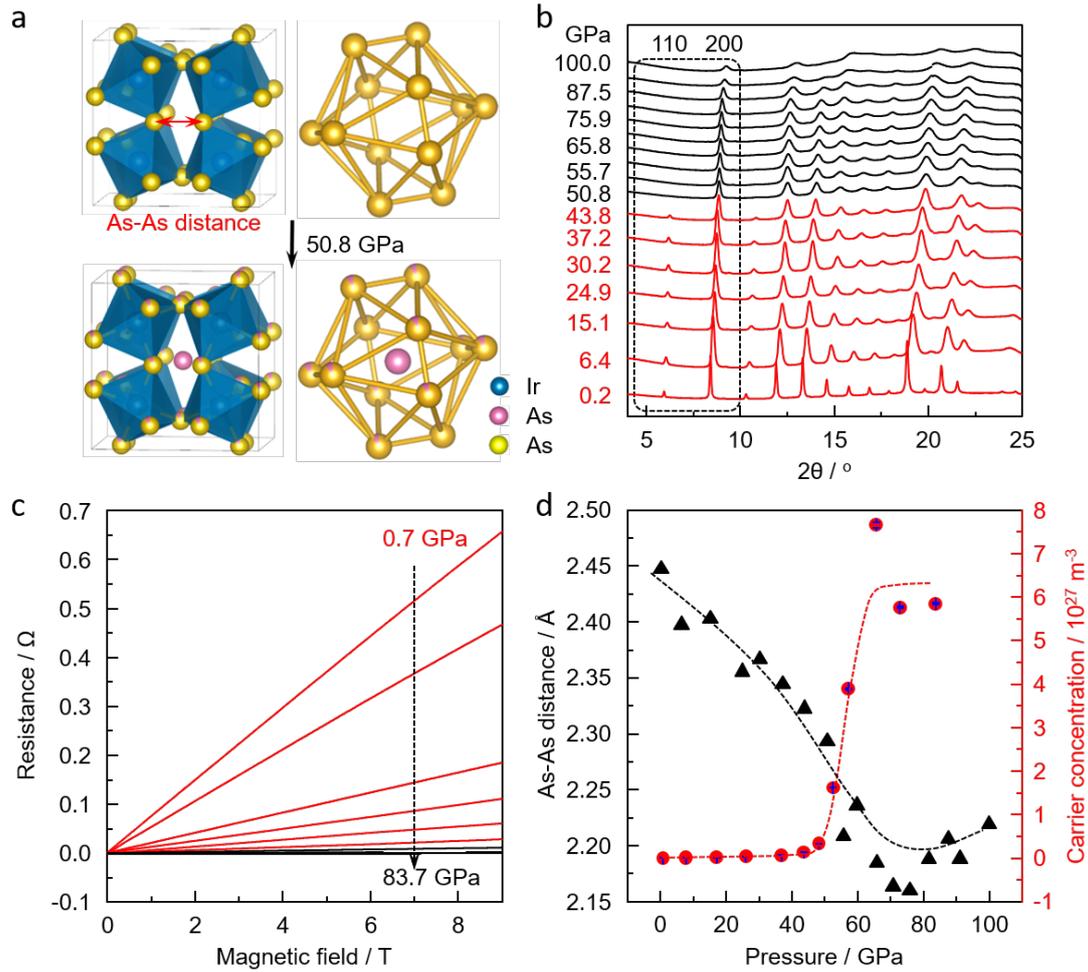

**Figure 1.** (a) Schematic of the crystal structures of IrAs$_3$ before and after atom insertion. (b) XRD patterns of IrAs$_3$ measured at room temperature with increasing of external pressure up to 100 GPa. The X-ray diffraction wavelength $\lambda$ is 0.6199 Å. Red and black curves are used to distinguish the structure transformation at 50.8 GPa. (c) Hall resistance of IrAs$_3$ as a function of magnetic field under various pressures. (d) Extracted As-As distance (from 1b) and carrier concentration (from 1c) under external pressures from 0.2 to 100 GPa.

The framework of skutterudites consists of eight tilted octahedra MX$_6$ per unit cell (M = Ir, X = As or P), with vertex sharing at X sites to enclose two icosahedral cages (Figure 1a). To investigate the influence of external pressure upon the crystal structure, we performed *in-situ* synchrotron X-ray diffraction (XRD) on IrAs$_3$ under various pressures from ambient condition to 100 GPa. As shown in Figure 1b, the diffraction peaks in the low-pressure range systematically shift to higher angles but share a similar pattern as the ambient IrAs$_3$ with the space group of *Im$\bar{3}$*. Above 50.8 GPa, 110 Bragg peak suddenly vanishes, while the other peaks keep shifting monotonically toward high angles. Accompanying the peak evolution, the volume of the unit cell also presents a discontinuity at the same critical pressure (Figure S2). The disappearance of the 110 peak within an identical space group hints at the possible insertion of atoms inside the cage. We simulate the synchrotron diffraction pattern of raw IrAs$_3$, Ir- and As-filled skutterudites in Figure S3. Because of the cancellation of structure factors of the caged and structural As, the intensity of the 110 peak almost vanishes, agreeing well with the experimental observation. This systematic distinction of the diffraction peaks cannot be preferred orientation according to our pressure-dependent powder X-ray diffraction pattern, where the uniform intensity of the diffraction circles can be seen (Figure S4). Similar behavior has been observed for IrP$_3$ (Figure S5). The enthalpy calculation (Figure S6) reveals that the formation energy of Ir filling of IrAs$_3$ is significantly higher than that of As filling, supporting the insertion of As inside the cage under high pressure.

Another direct evidence of the insertion of As comes from the pressure-dependent Hall measurements. Figure 1c shows selective Hall resistance curves $R_{xy}$(H) under various pressures. At 0.7 GPa, the Hall resistance curve exhibits a linear feature with a positive slope, indicating a hole-dominated feature of the electrical transport and in agreement with the data measured at ambient pressure. Unlike all the reported electron-doped skutterudites, the slope of the Hall resistance becomes much smaller as the external pressure increases, indicating the prominent hole doping effect. We extracted the carrier concentration and summarized it in Figure 1d, together with the As-As distance evolution. A prominent feature is the jump of the hole concentration near the critical pressure of 50.8 GPa. This is coincident with the slope change of the As-As distance. The high-pressure Hall resistance of IrP$_3$ is shown in Figure S7.

Figure S8 shows the band structures of Ir$_4$As$_{11.88}$ (1% As vacancies) and pristine IrAs$_3$, demonstrating a prominent hole doping effect. Given the electronegativity of As (2.18) and Ir (2.20), As counter-intuitively donates electrons to Ir in IrAs$_3$. Because of the large distance of 3 Å between the caged As and its surrounding 12 structural As, squeezing As inside the cage has an equivalent hole doping effect as moving it to an infinite distance. Namely, hole doping in the current system is accomplished by moving the position of positively charged As. It is worth noting that carrier doping is often accomplished through element substitution, intercalation, gate modulation, or chemical etching. The above findings represent a novel carrier doping strategy by rearranging the atom to a substantially isolated Wyckoff position under high pressure. Similar phenomena can be anticipated for other clathrates or caged materials under pressure.

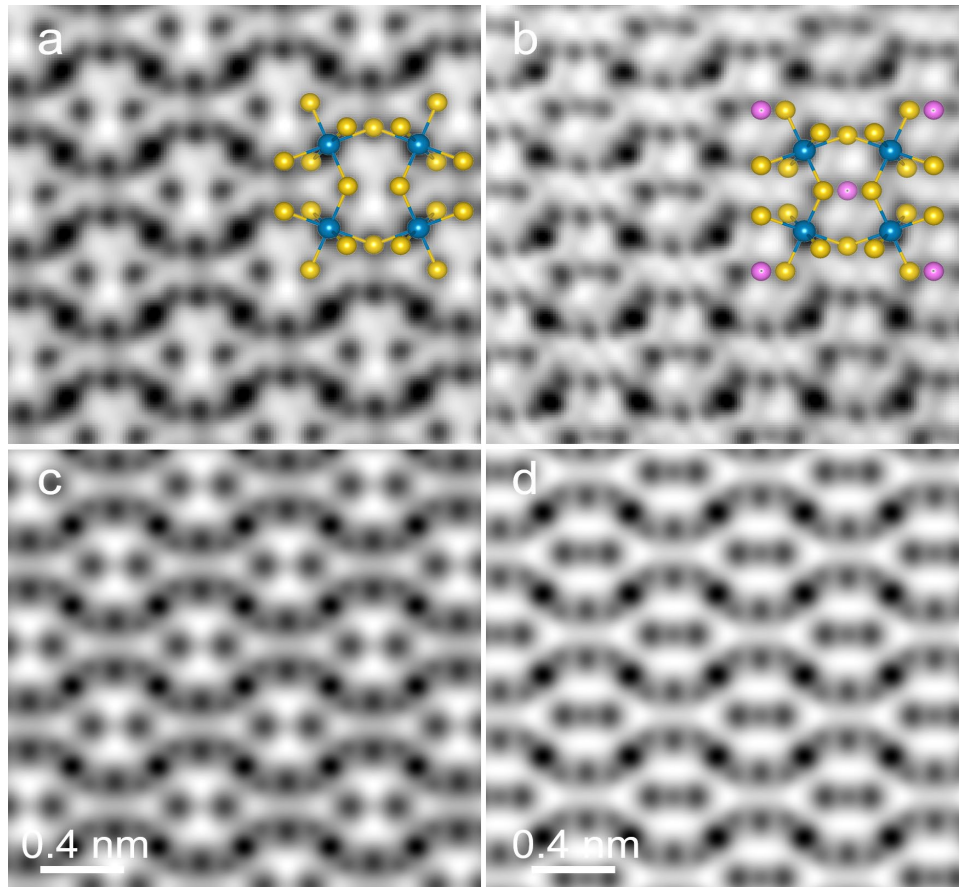

**Figure 2.** (a) Angular bright-field (ABF) image of pristine $IrAs_3$ and corresponding simulated ABF image (c). (b) ABF image of $As_\delta Ir_4 As_{12-\delta}$ (recovered from 65 GPa) and corresponding simulated ABF image (d). Their respective crystal structures are superimposed.

To visualize the inserted As atoms in real space, we investigate the atomic structure of pristine and high-pressure treated $IrAs_3$. Compared to the pristine $IrAs_3$ (Figure 2a), additional As atoms in the cage can be easily distinguished (Figure 2b). Figures 2c,d illustrates their simulated ABF patterns. As shown in Figure S9, partial As sites have a lower intensity than the remaining As on the cage, implying that As is self-inserted under high pressure. Judging from the intensity of the filled As with the reference of its nearby atomic columns (Figures 2b and S9), the filling rate of the center As at 65 GPa is about 60%. We use the notation of As$_\delta$Ir$_4$As$_{12-\delta}$ to emphasize the As atom inside the cage.

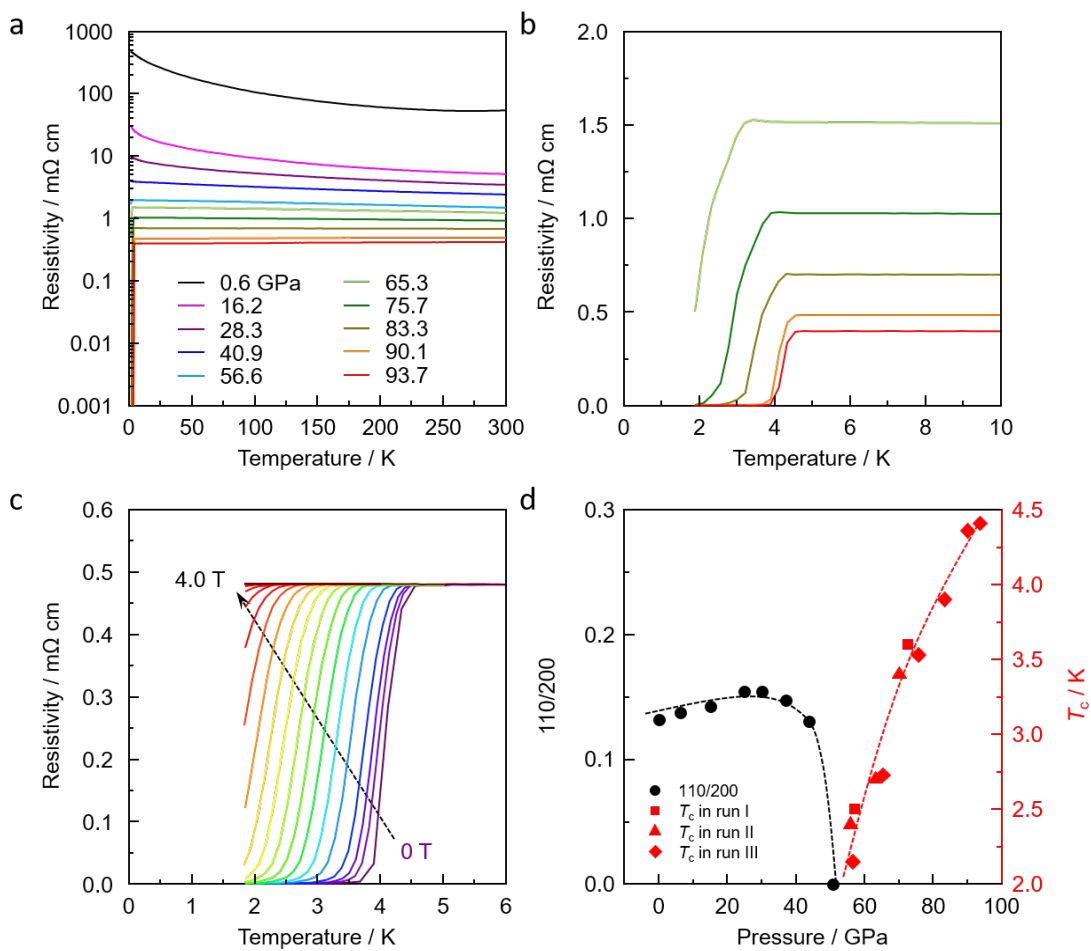

**Figure 3.** (a) Electrical resistivity of IrAs$_3$ as a function of temperature for pressures in run III. (b) Temperature-dependent resistivity of IrAs$_3$ in the vicinity of the superconducting transition. (c) Temperature dependence of resistivity under different magnetic fields for IrAs$_3$ at 90.1 GPa. The superconducting $T_c$ is gradually suppressed with increasing field. (d) Pressure dependence of the XRD peak 110/200 intensity ratio (from Figure 1b) and superconducting transition temperatures $T_c$ (from Figure 3b and Figure S10a) for IrAs$_3$ in different runs.

As the only hole-doped counterpart of the numerous electron-doped skutterudites discovered so far, it is intriguing to investigate the transport properties of hole-doped IrAs$_3$. We measured the electrical resistivity $\rho$(T) of IrAs$_3$ at various pressures. Figure 3a shows the typical $\rho$(T) curves for pressures up to 93.7 GPa. We find that with increasing pressure, it gradually changes from the semiconducting to a metallic behavior, followed by a small drop of $\rho$ that is observed at the lowest measured temperature ($T_{min}$ = 1.8 K). With further increasing pressure, zero resistivity is achieved at low temperatures for $P \geq 75.7$ GPa, indicating the emergence of superconductivity (Figure 3b). The superconducting $T_c$ increases dramatically with pressure and a maximum and unsaturated $T_c$ of 4.4 K is obtained within the limit of our research. The value of $\mu_0H_{c2}$(T) was estimated to be 3.4 T at 90.1 GPa (Figure 3c and Figure S10b), which yields a Ginzburg–Landau coherence length $\xi_{GL}$(0) of 9.85 nm. We noticed an interesting phenomenon in the decompressing process. $T_c^{onset}$ increases gradually with external pressure from 2.7 K at 40.9 GPa to 3.2 K at 56.6 GPa and then to 4.1 K at 75.7 GPa (Figure S11a), but it remains nearly constant at 4.1 K as the pressure decreases (Figure S11b).

The measurements on different samples of IrAs$_3$ from three independent runs provide consistent and reproducible results (Figure S10a), confirming the intrinsic superconductivity under pressure. The pressure dependence of the critical temperature for IrAs$_3$ is summarized in Figure 3d. Here we also use the normalized peak ratio of 110/200 to indicate the insertion of the As. Although a continuous lattice shrinkage is observed in the $a$-axis (Figure S2c) and the As-As distance, the ratio of 110/200 is almost invariant but suddenly drops to zero at 50.8 GPa, indicating the insertion of the As atom into the in-cage site. The coincidence of the 110 peak suppression and the $T_c$ emergence suggests a close correlation of superconductivity and As atom insertion. A similar phase diagram can also be obtained for IrP$_3$, with a maximum $T_c$ of 4.8 K at 98.7 GPa (Figures S12 and S13).

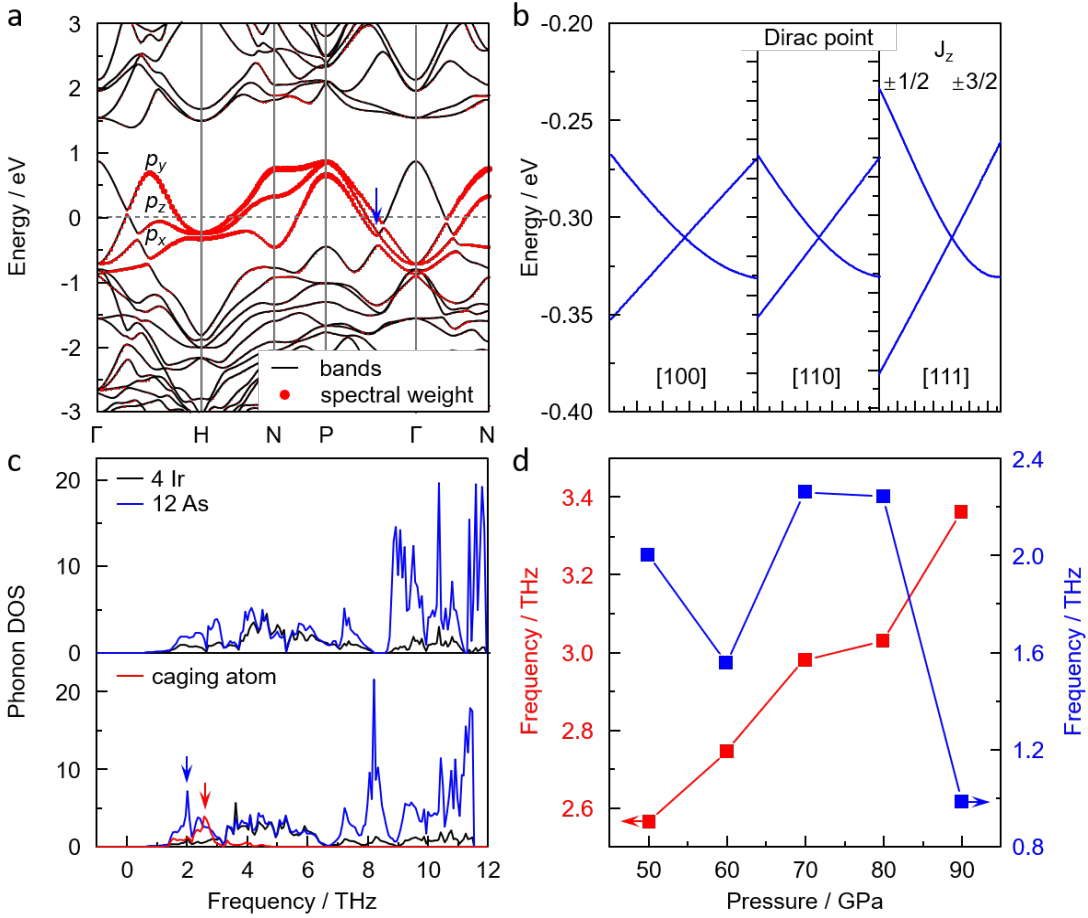

**Figure 4.** (a) Inserting As atoms into the cage generates three isolated bands with orbital characters $p_x$, $p_y$ and $p_z$ near the Fermi level, as indicated by the red color. The shown DFT band structure includes spin-orbit coupling (SOC). (b) Dispersion of the Dirac point indicated by the blue arrow in (a) along three directions, clearly showing the linear dispersion. (c) Phonon DOS of IrAs$_3$ before (up) and after (down) the self-insertion of the As atoms. (d) Evolution of the rattling mode (red arrow shown in (c)) together with the frequency of a newly generated phonon peak (blue arrow shown in (c)) with the variation of external pressure.

For electron-doped skutterudites, the caged cations generally donate the outer-shell electrons and do not participate in the bands near the $E_F$. Take BaIr$_4$P$_{12}$ and BaIr$_4$As$_{12}$ for example, where charge transfer of Ba pushes $E_F$ into the conduction bands, leaving its empty $s$-orbital far above $E_F$. Meanwhile, the conduction bands are composed of the hybridization of $5d$ orbitals from four Ir atoms with $p$ orbitals from 12 structural As atoms per primary cell, making the conducting bands extremely complex for further investigation. Numerous discussions on the superconductivity of skutterudites have focused only on a rough description of the partial density of states (PDOS) without a rigorous analysis of their electron components[5, 12, 27]. In contrast to guest cations in conventional filled skutterudites, the guest As atoms inserted into the cage of $As_\delta Ir_4 As_{12-\delta}$ tend to weaken the electron donation capability (Figure 4a), and thus will induce hole doping. A similar feature can be found in $AsIr_4As_{11}$ (Figure S14), where the crossing point shifts above Fermi level. The downwards shifting of the $E_F$ into the valance bands well supports our Hall measurements of hole doping (Figures 1 and S7).

Moreover, the As-$p$ orbital valence band of $IrAs_3$ is simple, implying that hole doping can generate clean Fermi surfaces. The Fermi level crosses the valence band, clearly demonstrating the hole doping effect of the guest atom. Most prominently, three isolated bands appear near the Fermi level and they are mainly attributed to $p$ orbitals of guest As atoms according to our analysis. These bands can be well described by a three-band tight-binding model including $p_x$, $p_y$ and $p_z$ orbitals centered at the guest As site (see details in SI and Figure S1). The narrow bandwidth originates from the weak hybridization within orbitals and leads to a non-negligible correlation effect. The cubic hole Fermi surface (FS) around Γ point is attributed to the valence band and the other FSs are from the guest As atoms. Furthermore, $X_\delta Ir_4 X_{12-\delta}$ is a topological material with a Dirac point formed by a crossing of the valence band with isolated bands from the guest As atoms shown in Figure 4b. The topological band structure intertwined with superconductivity may lead to topological superconductivity.

The filled caged materials are known for their relatively isolated phonons, which lead to "rattling" modes.[28] Previous works on the electron-doped skutterudites show that the Debye temperature depends mainly on the species forming the cage, while the Einstein temperature is found to roughly correspond to the low-energy optical modes of the guest atom. The inharmonic vibration of the caging atom and its interaction with electrons are generally believed to be responsible for the appearance of superconductivity in the electron-doped skutterudites.[5] The density of states of phonons for $IrAs_3$ and $X_\delta Ir_4 X_{12-\delta}$ are displayed in Figure 4c. We find that phonon modes from guest As atoms are localized around a low frequency, e.g. 2.5 THz at 50 GPa, in contrast to that of the undoped IrAs$_3$, where the phonons from the structural As are mainly located at a much higher-frequency region of about 7~12 THz. It is worth pointing out that the phonon contribution from one caging atom at 2.5 THz is comparable to the sum of the remaining 12 As atoms forming the cage. We also notice the emergence of an additional peak at 2 THz (blue arrow in Figure 4c) from the structural As, which may be caused by the phonon drag from the guest atom. To clarify the role of the guest atom and the concomitant superconductivity, we trace the evolution of the rattling mode (red arrow in Figure 4c) together with the newly generated peak (blue arrow in Figure 4c). As summarized in Figure 4d, the peak position of the rattling phonon continuously shifts to a higher value, agreeing well with the $T_c$ enhancement under pressure. On the contrary, the peak corresponding to the structural As behaves quite irregularly to external pressure. In the framework of BCS (electron-phonon interaction), the phonon frequency is correlated inversely with the $T_c$. However, in quite a few superconductors including cuprates and several others[29-32], their $T_c$s increase with the increase of phonon frequency. The concomitance of the rattling phonon and superconductivity implies their intimate relationship, although the underlying mechanism requires further exploration.

**CONCLUSIONS**

In summary, we successfully synthesized hole-doped skutterudites IrX$_3$ (X = P and As) by self-insertion of X atoms under high pressure. A superconducting state emerged with a maximum $T_c$ of 4.8 K, which can be attributed to the coupling of the rattling phonon mode with the electrons of three isolated $p$ bands at the Fermi level. The analyzable and simple band structures near $E_F$ may provide a unique opportunity to deepen the understanding of the superconductivity in skutterudites. The current findings present a novel carrier doping strategy by rearranging the atomic position without altering the space group or introducing impurities.

## ASSOCIATED CONTENT

**Supporting Information**. Experimental section, diffraction patterns, typical Rietveld refinement, electrical resistivity and calculated formation energy are supplied as Supporting Information.

## ACKNOWLEDGMENT

This work was supported by the National Key R&D Program of China (Grant No. 2018YFA0704300), the National Natural Science Foundation of China (Grant No. U1932217, 11974246 and 12004252), the Natural Science Foundation of Shanghai (Grant No. 19ZR1477300), the Science and Technology Commission of Shanghai Municipality (19JC1413900) and Shanghai Science and Technology Plan (Grant No. 21DZ2260400). The research at TIT was supported by MEXT Elements Strategy Initiative to form Core Research Center. (No. JPMXP0112101001). The authors thank the support from C$\hbar$EM (02161943) and Analytical Instrumentation Center (SPST-AIC10112914), SPST, ShanghaiTech University. The authors thank the staffs from BL15U1 at Shanghai Synchrotron Radiation Facility for assistance during data collection.